\documentclass[fleqn,usenatbib]{mnras}

\usepackage[T1]{fontenc}

\DeclareRobustCommand{\VAN}[3]{#2}
\let\VANthebibliography\thebibliography
\def\thebibliography{\DeclareRobustCommand{\VAN}[3]{##3}\VANthebibliography}

\usepackage{graphicx}	
\usepackage{amsmath}	
\usepackage{amssymb}	

\newcommand{\RNum}[1]{\uppercase\expandafter{\romannumeral #1\relax}}
\newcommand{\hii}{H\,\RNum{2}}
\newcommand{\hi}{H\,\RNum{1}}
\newcommand{\nii}{[N\,\RNum{2}]}
\newcommand{\sii}{[S\,\RNum{2}]}
\newcommand{\oiii}{[O\,\RNum{3}]}

\usepackage{newtxtext,newtxmath}

\title[M101 diffuse ionised gas]{Implications on star-formation-rate indicators from \hii{} regions and diffuse ionised gas in the M101 Group}

\author[A. E. Watkins et al.]{
A. E. Watkins,$^{1}$\thanks{E-mail: a.emery.watkins@gmail.com (AEW)}
J. C. Mihos,$^{2}$
P. Harding,$^{2}$
and R. Garner, III$^{3}$
\\
$^{1}$Centre for Astrophysics Research, School of Physics, Astronomy and Mathematics, University of Hertfordshire, Hatfield AL10 9AB, UK\\
$^{2}$Department of Astronomy, Case Western Reserve University, Cleveland OH 44106, USA\\
$^{3}$George P. and Cynthia W. Mitchell Institute for Fundamental Physics \& Astronomy, Texas A\&M University, College Station, TX 77843, USA
}

\date{Accepted XXX. Received YYY; in original form ZZZ}

\pubyear{2023}

\begin{document}
\label{firstpage}
\pagerange{\pageref{firstpage}---\pageref{lastpage}}
\maketitle

\begin{abstract}
We examine the connection between diffuse ionised gas (DIG), \hii{} regions, and field O and B stars in the nearby spiral M101 and its dwarf companion NGC~5474 using ultra-deep H$\alpha$ narrow-band imaging and archival GALEX UV imaging. We find a strong correlation between DIG H$\alpha$ surface brightness and the incident ionising flux leaked from the nearby \hii{} regions, which we reproduce well using simple Cloudy simulations. While we also find a strong correlation between H$\alpha$ and co-spatial FUV surface brightness in DIG, the extinction-corrected integrated UV colours in these regions imply stellar populations too old to produce the necessary ionising photon flux. Combined, this suggests that \hii{} region leakage, not field OB stars, is the primary source of DIG in the M101 Group. Corroborating this interpretation, we find systematic disagreement between the H$\alpha$- and FUV-derived star formation rates (SFRs) in the DIG, with SFR$_{{\rm H}\alpha} < $SFR$_{\rm FUV}$ everywhere. Within \hii{} regions, we find a constant SFR ratio of $0.44$ to a limit of $\sim10^{-5}$ M$_{\odot}$~yr$^{-1}$. This result is in tension with other studies of star formation in spiral galaxies, which typically show a declining SFR$_{{\rm H}\alpha}/$SFR$_{\rm FUV}$ ratio at low SFR. We reproduce such trends only when considering spatially averaged photometry that mixes \hii{} regions, DIG, and regions lacking H$\alpha$ entirely, suggesting that the declining trends found in other galaxies may result purely from the relative fraction of diffuse flux, leaky compact \hii{} regions, and non-ionising FUV-emitting stellar populations in different regions within the galaxy.
\end{abstract}

\begin{keywords}
ISM: HII regions --- ISM: clouds --- ISM: evolution --- galaxies: star formation --- ultraviolet: galaxies
\end{keywords}



\section{Introduction}
\label{sec:intro}

The interstellar medium (ISM) comprises the fuel behind star formation in galaxies.  While the stars themselves form from the cold ISM, primarily molecular hydrogen, feedback from this star formation in the form of supernovae, stellar winds, and high-energy photons ensures that much of the ISM exists in a high-temperature, ionised state \citep{mckee77, madsen06, haffner09}.  The ionised ISM thus contains a ledger of the ionising potential of galaxies, a record of which is critical for understanding both the impact of baryonic feedback on galaxy evolution, a necessary constraint on structural cosmological parameters \citep[e.g.,][]{jing06, vandaalen11, chisari18}, and the early evolution of galaxies during the epoch of reionisation, an era now increasingly accessible with the advent of JWST \citep[e.g.,][]{windhorst23}.

Of that ionised gas, an important component is morphologically diffuse, and so at a glance appears unassociated with any specific ionisation source.  \citet{hoyle63} first proposed the existence of this diffuse ionised gas layer in the Milky Way (MW) based on the detection of a free-free absorption signature in the Galactic synchrotron background by \citet{reber56} and \citet{ellis62}.  \citet{reynolds73a} eventually detected this layer directly in H$\alpha$ and H$\beta$ emission, and later observations with the Wisconsin H$\alpha$ Mapper \citep[WHAM;][]{haffner03} found that faint H$\alpha$ emission is ubiquitous in the northern sky to a surface brightness of $I_{{\rm H}\alpha}=0.1$~R ($\sim 5.7\times10^{-19}$ erg s$^{-1}$ cm$^{-2}$ arcsec$^{-2}$).  \citet{gaustad01} found similar results in the Southern hemisphere to slightly less sensitivity (0.5~R).  \citet{dettmar90} and \citet{rand90} first identified extragalactic DIG in the edge-on disk galaxy NGC~891, above and below the disk plane, and subsequently it was found in interarm regions in lower inclination disk and irregular galaxies \citep{hunter90, walterbos92, ferguson96}.  It became clear that this diffuse ionised gas \citep[DIG, dubbed the warm ionised medium by][]{mckee77} is ubiquitous in star-forming galaxies.  It occasionally even appears far outside its putative host \citep[e.g.,][]{devine99, lehnert99, keel12, watkins18}.

DIG properties are deeply connected to the structure of the ISM \citep[e.g.,][]{wood05, seon09}, hence constraining its ionisation source is critical for understanding how such radiation propagates through gaseous media.  From the earliest investigations, it was clear that photoionisation must be an important such source.  For example, the power necessary to ionise DIG is comparable to the total power injected by luminous young stars and supernovae, both in the MW \citep[e.g.,][]{reynolds90} and elsewhere \citep[e.g.,][]{ferguson96}.  In most DIG, little extra heating beyond photoionisation is required to model its observed spectra \citep{domgorgen94, mathis00}.

The path those ionising photons take, however, is less straight-forward. DIG spectra differ from that of the more compact \hii{} regions, being often relatively elevated in \nii{}6549,6584\AA \ (hereafter, \nii{}), \sii{}6716,6731\AA \ (hereafter, \sii{}), [O\,\RNum{1}]6300\AA, and other low-ionisation emission lines.  Typically, these line strengths increase with decreasing H$\alpha$ surface brightness \citep[e.g.,][]{madsen06, haffner09, hill14} and with height above the midplane \citep[e.g.,][]{rand98, haffner99, otte01, miller03, levy19}, albeit with wide variability.  Photoionisation simulations demonstrate that this can be achieved via leakage of ionising photons from \hii{} regions: because photons with energies near the ionisation potential of hydrogen (13.6~eV) are preferentially absorbed in a neutral medium, the ionising spectrum of the Lyman continuum (LyC) photons which propagate into the diffuse ISM tends to be harder than that found within \hii{} regions \citep[e.g.,][]{wood04}, increasing the kinetic energy of electrons and thus gas temperature.  Lines such as \nii{} and \sii{} are predominantly collisional \citep{osterbrock06}, thus elevated \nii{}/H$\alpha$ and \sii{}/H$\alpha$ line ratios imply higher gas temperatures.

Yet \oiii{}5007\AA, which has a much higher ionising potential ($\sim35$~eV), sometimes also increases with height beyond what is expected from a hardening LyC spectrum alone \citep[e.g.,][]{rand98, collins01}, suggesting some additional ionising component is necessary to fully explain DIG spectra.  Such additional proposals range widely, from post-ABG stars in the stellar halo or thick disk \citep[also known as hot low-mass evolved stars, or HOLMES; e.g.,][]{wood04, floresfajardo11, rautio22}, to shock ionisation from supernova or AGN feedback \citep[e.g.,][]{dopita95, simpson07, ho14}, to magnetic recombination \citep{raymond92, lazarian20}.  Most likely, many such mechanisms contribute to DIG ionisation in different amounts depending on local phenomena, such as the creation of superbubbles \citep{madsen06, rautio22}, so the exact fractional contribution of each is still a matter of debate.

Even the photoionisation budget is not completely clear, however, as O and B stars outside of \hii{} regions likely contribute to DIG ionisation to some extent \citep[possibly nearly 40\%;][]{hoopes00, hoopes01}.  Such field O and B stars have been identified in the MW and its satellites \citep[e.g.,][]{gies87, oey04, lamb13}, and many star-forming galaxies also host substantial extended diffuse FUV components \citep[][]{gildepaz05, thilker05b, thilker07}.  Some orphan O and B stars are host to their own spherical, ghostly \hii{} regions \citep[e.g.,][]{oey13}, implying that they do ionise their local ISM and thus can contribute to DIG.

However, it remains unclear where these stars originate, and therefore how their impact might vary as a function of environment.  While some may form in the field directly, most others likely formed within clusters and later drifted \citep["walk-away" stars;][]{demink12, renzo19} or were jettisoned \citep["run-away" stars;][]{blaauw61} to their current locations \citep[e.g.,][]{oey04, dewit05, lamb10, vargassalazar20}.  Depending on their velocities, these stars may not stray far from their birth clusters before dying.

If \hii{} region leakage is the primary source of DIG photoionisation, one might expect a weak correlation between H$\alpha$ and FUV flux in DIG regions, but a strong correlation between the estimated incident ionising flux from a galaxy's \hii{} regions and DIG surface brightness \citep[e.g.,][]{zurita02, seon09, belfiore22}.  DIG would also be predominantly found surrounding \hii{} region complexes, as geometric dilution and neutral ISM absorption would prevent gas ionisation elsewhere \citep[save extraplanar DIG, where the plane-parallel approximation is more appropriate and most of the ISM is ionised; e.g.,][]{berkhuijsen06, floresfajardo11}.  If, on the other hand, in-situ field O and B stars are the primary source, we would see the inverse behavior in the correlations, and the spatial distribution of DIG would depend on the origins, lifespans, and velocities of the ionising stars.   Contribution from shock ionisation or AGN would likely be localized to supernova remnants and galaxy cores, respectively, but would be difficult to isolate without additional diagnostic lines, while HOLMES contribution would be found primarily where old FUV-weak stellar populations dominate, such as the bulge or stellar halo \citep[e.g.,][]{lacerda18}.

The connection between diffuse H$\alpha$ and diffuse FUV is also mystified somewhat by a well-known discrepancy between H$\alpha$- and FUV-derived star formation rates (SFRs) in low surface brightness (LSB) regions.  Both in the FUV-emitting outer disks of massive galaxies \citep{goddard10, byun21} and in dwarf and LSB galaxies \citep[e.g.,][]{lee09, meurer09, lee16}, the ratio between H$\alpha$-derived and FUV-derived SFRs is depressed, often to below 50\% \citep[but see][]{bell01}.  Proposed mechanisms behind this observation in the low-density regime include changes in the stellar initial mass function \citep[e.g.,][]{meurer09, pflamm09}, higher LyC escape fraction in low-mass systems \citep[e.g.,][]{relano12}, and less efficient star formation resulting in more sporadic star formation histories \citep[SFHs; e.g.,][]{sullivan04, weisz12, emami19}.  Thus, understanding the origins of DIG is key to the proper utilisation of H$\alpha$ emission as an SFR indicator on large spatial scales, and may help illuminate the fundamental physics behind star-formation as a whole.

To help provide more constraints on the ionisation sources of DIG, we explore the diffuse H$\alpha$ and FUV emission in the M101 Group, a local \citep[$D=6.9$~Mpc;][]{matheson12} loose association of galaxies.  The group is rather sparse, containing only the massive \citep[$\log(\mathcal{M}_{*})=10.6$;][]{munozmateos15} face-on spiral M101 (NGC~5457), its lower mass \citep[$\log(\mathcal{M}_{*})=9.1$;][]{munozmateos15} companion NGC~5474, the irregular star-forming dwarf NGC~5477, and a handful of much fainter satellite candidates, most only recently identified \citep{muller17}.  Indeed, a recent survey using the Hubble Space Telescope found that M101's satellite population is very sparse, with roughly half the number of low-mass companions as the MW to a limit of $M_{V}=-7.7$ \citep{bennet20}.  However, this low group mass, and consequent low velocity dispersion, should allow for more impactful tidal interactions between the group members \citep{negroponte83}.  Integrated light \citep{mihos13} and resolved stellar \citep{mihos18} photometry of M101's outer disk has illustrated the impact of this on the star formation history (SFH): a burst of star formation which peaked 300---400~Myr ago in M101's outer disk.  Follow-up simulations demonstrate that the most massive companion, NGC~5474, is the most likely culprit \citep{linden22}.

The group's well-characterized SFH thus makes it a useful target for studying the origins of the DIG, as it allows one to marginalize SFH as a possible parameter when interpreting H$\alpha$/FUV SFR ratios.  We thus explore the relationship between \hii{} region emission and the DIG surface brightness, as well as that between FUV and H$\alpha$ emission in the DIG, within the M101 group's most massive members, using archival GaLaxy Evolution EXplorer \citep[GALEX;][]{martin05} ultraviolet imaging and our own ultra-deep H$\alpha$ narrow-band imaging done with the Burrell Schmidt Telescope.  We give a brief summary of our observations and archival data in Sec.~\ref{sec:obsdat} and Table~\ref{tab:obs}.  We describe our methodology behind photometry of DIG and \hii{} regions in Sec.~\ref{sec:methods}, focusing on measurement and corrections to systematics such as extinction.  We present scaling relations derived from our systematics-corrected H$\alpha$ and FUV measurements in Sec.~\ref{sec:results}.  We discuss these results in Sec.~\ref{sec:discussion}, and finally provide a full summary in Sec.~\ref{sec:summary}.

\section{Observations summary}
\label{sec:obsdat}

\begin{table*}
    \centering
    \begin{tabular}{ccccccccc}
    \hline\hline
    Survey  &  Galaxy  &  Band  &  $\lambda_{\rm cen}$ &  $\Delta \lambda$  &  FWHM  &  t$_{\rm exp}$  &  RMS  & Reference \\
    (1)  &  (2)  &  (3)  &  (4)  &  (5)  &  (6)  &  (7)  &  (8)  &  (9)  \\
    \hline  \\
    BST$_{2009}$  &  Both  &  $B$  &  4107~\AA  &  1067~\AA  & 2.5\arcsec  &  $59\times1200$~s  & $5.40\times10^{-20}$ & \citet{mihos13} \\
    BST$_{2010}$  &  Both  &  $M$  &  5088~\AA  &  1207~\AA  & 2.2\arcsec  &  $60\times900$~s  & $3.25\times10^{-20}$ & \citet{mihos13} \\
    BST$_{2014}$  &  Both  &  H$\alpha$-on  &  6590~\AA  &  101~\AA  &  2\arcsec  &  $71\times1200$~s  &  $9.97\times10^{-18}$ & \citet{watkins17} \\
    BST$_{2014}$  &  Both  &  H$\alpha$-off  &  6726~\AA  &  104~\AA  &  2\arcsec  &  $71\times1200$~s  &  $9.53\times10^{-18}$ & \citet{watkins17} \\
    BST$_{2018}$  &  Both  &  H$\beta$-on  &  4875~\AA  &  82~\AA  &  2.3\arcsec &  $59\times1200$~s  & $6.72\times10^{-18}$ & \citet{garner21} \\
    BST$_{2018}$  &  Both  &  H$\beta$-off  &  4757~\AA  &  81~\AA  &  2.3\arcsec &  $55\times1200$~s  &  $6.80\times10^{-18}$ & \citet{garner21} \\
    \emph{GALEX} GI3\textunderscore 050008  &  M101  &  FUV  &  1530~\AA &  265~\AA  &  4.2\arcsec  &  13293.4~s  &  $2.20\times10^{-19}$ & \citet{leroy08} \\
    \emph{GALEX} GI3\textunderscore 050008  &  M101  &  NUV  &  2303~\AA &  768~\AA  &  4.9\arcsec  &  13293.4~s  &  $9.87\times10^{-20}$ & \citet{leroy08} \\
    \emph{GALEX} NGA &  NGC~5474  &  FUV  &  1530~\AA  &  265~\AA  &  4.2\arcsec  &  1610~s  &  $1.04\times10^{-18}$ & \citet{gildepaz07} \\
    \emph{GALEX} NGA &  NGC~5474  &  NUV  &  2303~\AA  &  768~\AA  &  4.9\arcsec  &  1610~s  &  $2.99\times10^{-19}$ & \citet{gildepaz07} \\
    \hline
    \end{tabular}
    \caption{Table of imaging data used.  The columns are: 1 $-$ name of survey or instrument from which data is taken; 2 $-$ galaxy covered by survey data given by $1$; 3 $-$ photometric band; 4 $-$ effective wavelength of photometric band; 5 $-$ width of photometric band;  6 $-$ PSF FWHM of image coadds; 7 $-$ total integration time on target galaxy; 8 $-$ background root-mean-square uncertainty (ergs s$^{-1}$ cm$^{-2}$ \AA$^{-1}$ for broadband, ergs s$^{-1}$ cm$^{-2}$ for narrow-band); 9 $-$ reference paper for observations.
    \vspace{0.1cm}}
    \label{tab:obs}
\end{table*}

We use imaging data from two different observatories for our study.  First, we use broadband and narrow-band images of M101 and NGC~5474 taken with the Burrell Schmidt Telescope (BST) at Kitt Peak National Observatory, a 0.6/0.9m telescope optimized for LSB imaging.  Broadband observations were taken in April of 2009 and April of 2010, in a modified Johnson $B$-band filter ($\sim200$\AA \ bluer than standard) and in Washington $M$, respectively \citep{mihos13}.  This broadband imaging is calibrated directly to Johnson B and V magnitudes using stars in the field surrounding M101; the details of the photometric calibration can be found in \citet{mihos13}.  Narrow-band observations were taken in April through June of 2014 \citep[H$\alpha$;][]{watkins17} and March through May of 2018 \citep[H$\beta$;][]{garner21}, using custom $\sim100$\AA-wide filters for both on- and off-band observations.  In addition, we used archival GALEX far-ultraviolet (FUV) and near-ultraviolet (NUV) imaging.  For M101, we used the deep imaging from the guest investigator program published in \citet{leroy08}, and for NGC~5474, we used the imaging from the Nearby Galaxies Atlas \citep[NGA;][]{gildepaz07} program.

We summarize the observations in Table~\ref{tab:obs}, including survey, photometric band, resolution on the image coadds, total integration times on-target, and pixel-to-pixel root-mean-square (RMS) uncertainty in the background (in physical flux units).  Details of the observation and data reduction strategies used for each set of observations can be found in the associated references (column 9).  The BST and GALEX pixel scales are 1.45 arcsec px$^{-1}$ and 1.5 arcsec px$^{-1}$, respectively.

The RMS uncertainty for the H$\alpha$ difference image is 7.97$\times 10^{-18}$ erg s$^{-1}$ cm$^{-2}$, slightly lower than in either narrow-band image individually.  This is because much of the background uncertainty in low-resolution imaging comes from unresolved sources, many of which subtract out when creating the difference image.

\section{Methods}
\label{sec:methods}

We present here our procedure for identifying, measuring the fluxes of, and applying photometric corrections to both \hii{} regions and DIG.  We required photometry of both diffuse emission and the more compact \hii{} regions first to separate the two components, and second to assess the impacts of both \hii{} region leakage and field O and B stars on DIG properties.

Our initial catalogue of \hii{} region candidate sources is also likely a mix of true \hii{} regions and interlopers.  Thus, we first apply physical corrections to all \hii{} region candidate source photometry, then make interloper cuts based on these physically-corrected parameters.

\subsection{Detection}
\label{ssec:detection}

\begin{figure}
    \centering
    \includegraphics[scale=1.0]{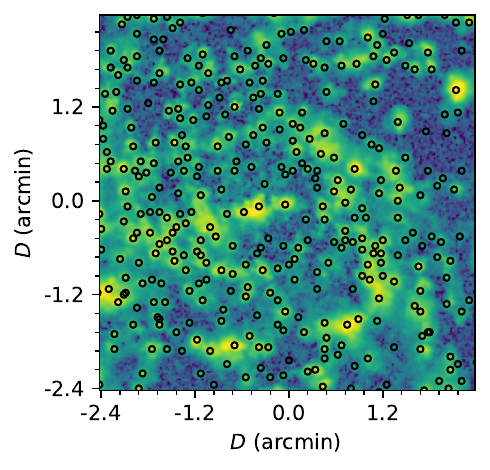}
    \caption{Demonstrating results of our point-like source identification algorithm for the central regions of M101.  Black circles are centred on the sources identified by re-centring the segmentation map produced by Sourcerer (see text), overlaid on a logarithmically scaled image of the BST H$\alpha$ difference image on which we performed the segmentation.  Axis labels are distance from the center of M101 in arcminutes.  At M101's distance, 1\arcmin$\sim 2$~kpc.}
    \label{fig:peaks}
\end{figure}

We identified both candidate \hii{} regions and DIG regions with the software Sourcerer \citep[formerly, MTObjects;][]{teeninga13, teeninga16, haigh21}.  Briefly, Sourcerer performs object detection and segmentation using a max tree algorithm \citep{salembier98}, which identifies local maxima in an image's flux distribution as leaves of a tree, and nodes as increasingly large connected regions of the image (with the root represented by the entire image).  Only those local maxima and connected nodes determined to be statistically significant against the local background are designated as detections, assuming the background has normally distributed noise.  Being based on local flux hierarchy, the segmentation Sourcerer performs makes it ideal for identifying embedded point sources, while being simultaneously sensitive to extremely LSB contiguous emission.  We ran the software on images cropped around each galaxy separately to ensure the background noise characteristics were not influenced by the variation in exposure counts in different parts of the coadd.

Without access to spectra of most of M101, particularly in its faint outskirts, disentangling DIG and \hii{} region flux can be an ambiguous task.  To simplify the process, in both galaxies we chose to label the point-like sources visible in our BST H$\alpha$ difference image as \hii{} region candidates and all other significant H$\alpha$ emission as DIG.  This is, to some extent, justified by our images' low resolution.  At M101's 6.9~Mpc distance, one arcsecond corresponds to $\sim33$~pc, while the distribution of extragalactic \hii{} region radii published by \citet{congiu23} shows a peak at roughly $\sim90$~pc.  Our H$\alpha$ narrow-band imaging has a FWHM$\sim 2$\arcsec (Table~\ref{tab:obs}), or 67~pc.  Therefore, typical \hii{} regions are barely resolved in our BST imaging, and are unresolved in FUV imaging given that instrument's much larger PSF (FWHM$\sim 4$\arcsec).

Even so, our decision to isolate point-like sources defines \hii{} regions as only the most compact and brightest parts of the ionised clouds, likely centred on young star clusters.  We recognize this definition differs from that used in many other extragalactic studies \citep[e.g.,][]{thilker00, erroz19, garner21}.  We consider its impact throughout our analysis.

While the software does segment images using a top-down approach, segmented regions containing point-like sources still typically contain flux from neighbouring diffuse pixels, and so are typically asymmetric.  To identify the centers of the point-like sources in each segmented region, we used either the coordinate of each region's brightest pixel, for regions with average surface brightness three times the RMS of the background local to each galaxy, or we used the flux-weighted centroid of the whole region, for sources fainter than this limit, to avoid noise peaks influencing the choice of central coordinate in LSB regions.  We show the coordinates of the point-like sources this method identified in the central regions of M101 in Fig.~\ref{fig:peaks}, overlaid on our H$\alpha$ difference image.

\begin{figure*}
    \centering
    \includegraphics[scale=1.0]{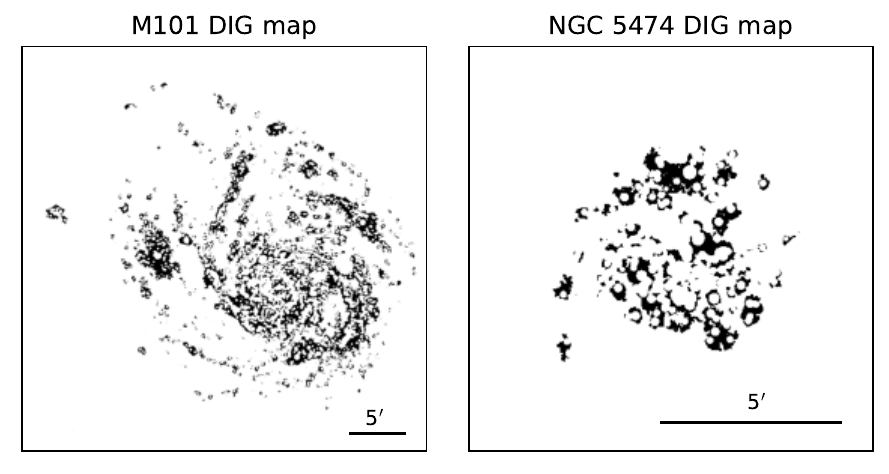}
    \caption{Pixel-to-pixel maps of diffuse ionised gas in M101 (left) and NGC~5474 (right).  All non-zero valued (black) pixels we consider DIG.  White pixels here are either masked sources or background.  We cleaned both maps by-hand of likely spurious detections, meaning those in the outskirts of either galaxy with patchy morphology and small angular size.  Black lines show 5\arcmin \ scale bars.  North is up and east is to the left in both panels.}
    \label{fig:digmap}
\end{figure*}

To separate DIG from \hii{} regions, we applied adaptive masks to each point-like source within the Sourcerer segmentation map.  For each source, we scaled the FUV PSF curve of growth to the source's total FUV flux to determine the radius at which the PSF surface brightness dropped a factor of five below our measured FUV limiting surface brightness (for M101 and NGC~5474: $\sim8.5 \times 10^{-20}$ and $1.2 \times 10^{-19}$ ergs s$^{-1}$ cm$^{-2}$ \AA$^{-1}$, respectively, $1\sigma$ on 100\arcsec$\times$100\arcsec \ scales).  We limited the mask radii between $2$~px$\leq R \leq 10$~px to avoid over- or under-masking; 10~px contains $> 95$\% of the total FUV PSF flux, hence even for extremely bright sources this aperture limit suggests only a maximum $\sim5$\% flux contamination for DIG pixels directly adjacent.

We initially assigned every unmasked pixel detected by Sourcerer as DIG.  However, even in the native resolution difference image, Sourcerer often identified correlated noise in the background as significant detections, leading to prominent background contamination in our DIG map.  Given our small galaxy sample, we opted simply to erase by-hand all Sourcerer detections at large radius with small size.  Comparing scaling relations measured using the initial and cleaned DIG maps shows that our by-hand cleaning removed only long, LSB tails mostly below each band's noise limit.  We show cleaned DIG maps for both galaxies in Fig.~\ref{fig:digmap}.  Any pixel in these maps with a non-zero value we consider a DIG detection.

Defined in this way, 90\% of our DIG pixels have surface brightnesses below $\log(\Sigma_{{\rm H}\alpha}) < 38.3$ erg s$^{-1}$ kpc$^{-2}$.  This is lower than the DIG--\hii{} region separation threshold proposed by \citet{zhang17} ($\log(\Sigma_{{\rm H}\alpha}) < 39$ erg s$^{-1}$ kpc$^{-2}$), though many of our DIG pixels have surface brightnesses exceeding their threshold.  Our choice to define embedded point sources as \hii{} regions is more comparable to the method employed by \citet{thilker00}, who define DIG--\hii{} region boundaries using the local gradient of the H$\alpha$ surface brightness, albeit ours is more stringent.

\subsection{Photometry}
\label{ssec:photometry}

To estimate the impact of \hii{} region flux leakage on surrounding DIG, we required \hii{} region flux estimates free of DIG contamination, and of contamination from neighbouring point-like sources.  We thus measured fluxes of each point-like source by masking all neighbouring sources, doing 2~px radius aperture photometry of each target (to avoid source crowding; see Fig.~\ref{fig:peaks}), then applying background and aperture corrections to estimate total \hii{} region fluxes.  We find that, with aperture corrections applied, 3~px aperture fluxes are consistent with 2~px aperture fluxes to within $\log(F_{\lambda})\pm0.1$ in all photometric bands; hence, our choice of aperture has no impact on the correlations we examine throughout this paper.

We measured local backgrounds as the sigma-clipped median fluxes of unmasked pixels within ring apertures centred on each source, with inner radii of 10~px and widths of 2~px (chosen to lie beyond the PSF 95\% flux radius in all photometric bands).  Subtracting these local backgrounds corrects both for DIG contamination and for line absorption from any underlying stellar population \citep[see][]{garner22}.  We used the aperture corrections published by \citet{morrissey07} for GALEX bands, and for the BST, we derived our own from our coadds, by stacking and normalizing point sources with signal-to-noise ratios $>100$ external to all resolved galaxies in the field.  We applied aperture corrections only to background-corrected fluxes to derive total fluxes for each source.

DIG photometry required measurements from both the diffuse H$\alpha$ and diffuse FUV emission, to compare DIG and field O and B star populations.  This comparison required the H$\alpha$ and FUV images to have the same pixel scale and resolution. Thus, to ensure our DIG segmentation maps matched between photometric bands, we first reprojected our H$\alpha$ on- and off-band images to the GALEX pixel scale (a tiny change, from 1.45\arcsec \ px$^{-1}$ to 1.5\arcsec \ px$^{-1}$) using the Astropy-affiliated package {\sc reproject} \citep[v0.8---specifically, {\sc reproject\textunderscore adaptive}, which we found best preserved both flux and surface brightness per pixel;][]{astropy:2022}.  We then convolved each narrow-band image with a normalized Gaussian kernel with $\sigma^{2} = \sigma_{\rm FUV}^{2} - \sigma_{\lambda}^{2}$, where $\sigma_{\lambda}$ refers to the standard deviation of our H$\alpha$ on- and off-band averaged coadd PSFs.  Differencing these convolved images resulted in reprojected difference images for each galaxy.

We generated final-generation DIG segmentation maps (Fig.~\ref{fig:digmap}) from these reprojected difference images (converted roughly to analog-to-digital units using the on- and off-band photometric zeropoints).  We measured DIG fluxes pixel-to-pixel using these maps.

\subsection{Photometric corrections}
\label{ssec:corrections}

Converting from raw to intrinsic fluxes required correcting for both extinction and for line contamination (\nii{} and \sii{}) in our narrow-band filters.  We describe how we estimated these corrections for both \hii{} regions and DIG in this section.

\subsubsection{\hii{} region corrections}
\label{sssec:hiicorrs}

\begin{figure*}
    \centering
    \includegraphics[scale=1.0]{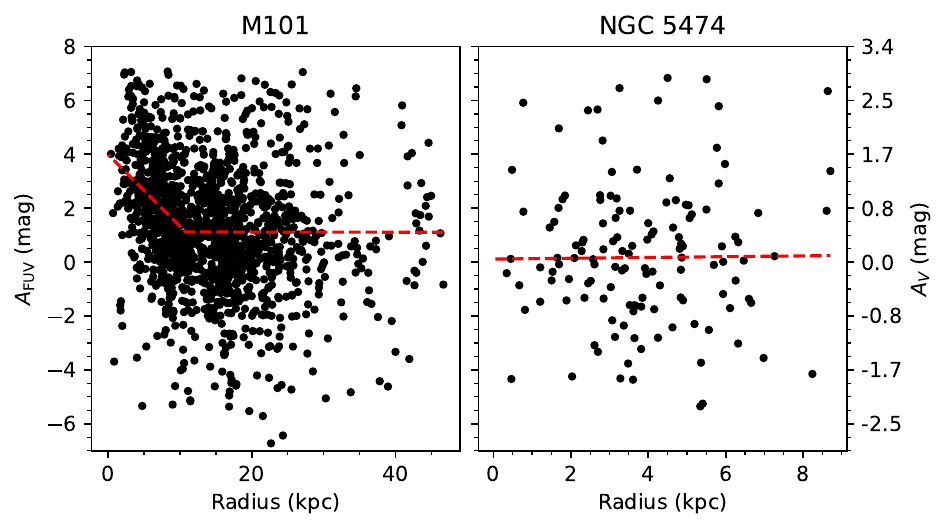}
    \caption{Demonstrating behavior of extinction internal to \hii{} regions in both galaxies.  Each panel shows the distribution of $A_{\rm FUV}$ as a function of radius, as derived from Balmer decrements.  The red dashed lines show the best-fit radial relation in each galaxy.  In M101, the gradient flattens beyond $\sim 11$~kpc, hence we assume a constant extinction beyond that radius, extrapolated from the last best-fit point in the inner relation.  NGC~5474 shows no radial extinction gradient and overall lower extinction than M101.  We also use these gradients to correct the DIG flux for extinction, albeit employing a different extinction law.}
    \label{fig:extcorr}
\end{figure*}

We first corrected all measured \hii{} region fluxes in all photometric bands for foreground MW extinction.  We derived extinction values $A_{\lambda}$ in all of our photometric bands using the Astropy-affiliated code {\sc dust\textunderscore extinction} \citep[v1.2;][]{astropy:2022}, selecting values of $A_{\lambda}/A_{V}$ from the average MW extinction curve from \citet{gordon09} and assuming $A_{V}=0.023$ \citep{schlafly11}.

To derive extinction internal to \hii{} regions in both galaxies, we used the Balmer decrement measured from our H$\alpha$ and H$\beta$ narrow-band images, assuming a theoretical value of $F_{{\rm H}\alpha}/F_{{\rm H}\beta}=2.86$\footnote{\citet{congiu23} recommend a theoretical value of $F_{{\rm H}\alpha}/F_{{\rm H}\beta}=3.03$ for DIG-corrected \hii{} region Balmer decrements in star-forming galaxies, but the intrinsic dispersion among measured Balmer decrements in our regions is high enough that the use of this value produced no noticeable change to the extinction gradients we derived in either galaxy.}.  We first corrected all H$\beta$ fluxes for internal stellar absorption following \citet{garner22}, by subtracting from each value $5$\AA \ EW of absorption.  This is the average absorption EW for \hii{} regions based on observed and model literature values \citep{gonzalez99, gavazzi04, moustakas06}.  Absorption in H$\alpha$ within \hii{} regions is typically negligible, and our local background subtraction removed any absorption from underlying older stellar populations within the galaxies. 

Fig.~\ref{fig:extcorr} shows the $A_{\rm FUV}$ gradient for \hii{} regions in both galaxies.  The region-to-region dispersion in measured Balmer decrements was quite high ($\sigma_{A_{V}}>0.6$~mag at any given radius within M101).  Some of this likely arises due to intrinsic variability in dust content and geometry among \hii{} regions; our lack of image resolution thus makes extinction measurements on any individual \hii{} region highly uncertain, even for bright regions with low photometric uncertainty.  Hence, for simplicity, we chose to apply global corrections as a function of radius, using linear fits between $A_{\lambda}$ and radius for both M101 and NGC~5474, ignoring all regions with $1 < F_{{\rm H}\alpha}/F_{{\rm H}\beta} < 8$ \citep[see][]{garner21}.  In M101, we found that extinction values flattened to $A_{V}\sim0.4\pm0.9$ (roughly $A_{\rm FUV}\sim1$; Fig.~\ref{fig:extcorr}) beyond $\sim300$\arcsec ($\sim10.8$~kpc), hence applied a constant extinction correction to all regions found beyond that radius by extrapolating from the last best-fit point interior to it.  NGC~5474 shows no clear radial gradient.  Red dashed lines in Fig.~\ref{fig:extcorr} show our best-fit extinction curves for both galaxies.

Our filter placement for the H$\alpha$ narrow-band imaging also included stellar continuum, H$\alpha$, and \nii{} emission in the on-band, and stellar continuum and \sii{} emission in the off-band.  We thus needed to estimate and correct for both of these emission lines in all derived H$\alpha$ fluxes.

We made these corrections for \hii{} regions in the manner described by \citet{garner22}, assuming a constant value of \sii{}/H$\alpha = 0.2$ and the radial \nii{}/\sii{} relation given by their Eq.~2, derived from \nii{} and \sii{} flux values in M101 published by \citet{croxall16}.  In NGC~5474, again following \citet{garner22}, we simply applied the average line correction factor for all \hii{} regions within M101 ($\sim0.95$).

\subsubsection{DIG corrections}
\label{sssec:digcorrs}

\begin{figure*}
    \centering
    \includegraphics[scale=1.0]{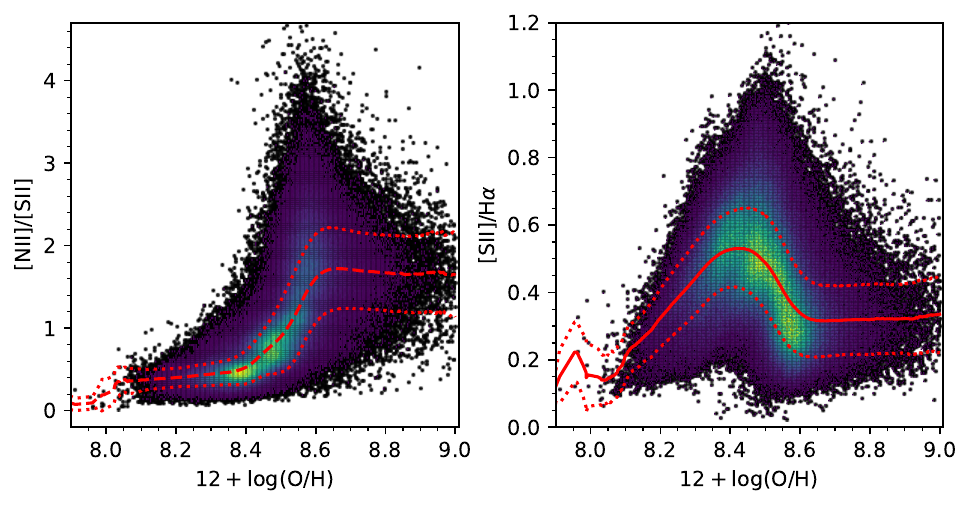}
    \caption{Distribution of \nii{}$/$\sii{} and \sii{}$/$H$\alpha$ fluxes in DIG spaxels among MUSE Atlas of Disks sample galaxies \citep[MAD;][]{erroz19}, as a function of gas-phase metallicity (using the MAD O3N2 calibration).  The red dashed lines show the running mean relations, while the red dotted lines show the running standard deviation about those means.  We use these relations to correct the DIG H$\alpha$ flux in M101 and NGC~5474 for \nii{} and \sii{} contamination.}
    \label{fig:niisii}
\end{figure*}

DIG represents a different physical system than \hii{} regions.  Hence, while we applied the same kinds of corrections to DIG, the forms of some of those corrections necessarily differed.  The exception is the MW extinction correction, which we applied in the same manner as for the \hii{} regions.

Internal extinction corrections here were less straight-forward.  In DIG-dominated regions, we found significant stellar absorption in H$\beta$ (and mild absorption in H$\alpha$) in both target galaxies, which is difficult to correct for without spectra.  Thus, we used publicly available data from the MUSE Atlas of Disks \citep[MAD;][]{erroz19}, in combination with the \hii{} region metallicities provided by \citet{croxall16}, to derive an empirical extinction correction for DIG based on our measured \hii{} region extinction gradients.  The MAD data includes, for 38 disk galaxies, line fluxes (including \nii{}, \sii{}, and H$\alpha$), measures of gas-phase metallicity, extinction derived from Balmer decrements, and also maps separating out DIG spaxels from \hii{} region spaxels using the method developed by \citet{blanc09}.

In all MAD galaxies, the median extinction among both \hii{} regions and DIG is roughly constant with radius (when normalized by the effective radius, though the scatter increases sharply toward the galaxy centers).  This results from the complex interplay between extinction, stellar mass surface density, and metallicity \citep{erroz19}.  The DIG extinction, however, shows similar trends to the \hii{} regions, but on average is lower by $A_{V}\sim0.1$.  Therefore, we assume the same applies for M101 and NGC~5474, and we use our same radial extinction corrections derived from the \hii{} region Balmer decrements to derive internal extinction in the DIG, but offset by $A_{V}=-0.1$.

DIG and diffuse FUV emission also arises from sources close to the disk plane and so are attenuated primarily by line-of-sight extra-planar dust.  Thus, for DIG and diffuse FUV, we use the stellar attenuation curve for low-inclination star-forming galaxies provided by \citet[Eq.~11, with $0.77 < b/a < 1$ coefficients from Table~1, assuming $R_{V}=3.64$, their extrapolated value from Table~2]{battisti17} to derive $A_{\lambda}$, rather than that of \citet{gordon09}.

Some stellar H$\alpha$ absorption was visible from our H$\alpha$ difference image, albeit mild.  Making use of Sourcerer's inability to detect negative-valued pixels, we estimated this by masking all Sourcerer detections in both galaxies, then using radial flux profiles of the unmasked flux to estimate corrections as a function of radius.  In M101, the correction in the inner 4~kpc \ was $\sim 4\times10^{-18}$ ergs s$^{-1}$ cm$^{-2}$ arcsec$^{-2}$ ($\sim13$\% of the median DIG H$\alpha$ surface brightness in the same region), decreasing linearly to zero by $\sim 540$\arcsec \ radius ($\sim 18$~kpc).  In NGC~5474, the correction in the inner 0.5~kpc was $\sim1.6\times10^{-17}$ ergs s$^{-1}$ cm$^{-2}$ arcsec$^{-2}$ ($\sim40$\% the median DIG H$\alpha$ surface brightness in that region), decreasing exponentially to zero by $\sim 110$\arcsec \ radius ($\sim 3.6$~kpc).  Hence, corrections in both galaxies were small.  However, these are lower-limits on the true absorption, as some DIG emission may be present in the regions with negative measured flux.

DIG typically shows enhanced \nii{}$/$H$\alpha$ and \sii{}$/$H$\alpha$ compared to \hii{} regions, hence requires different correction factors for these emission lines.  We derived these corrections for DIG by investigating the behavior of these lines within the MAD sample DIG spaxels.  The cleanest correlation lies between the \nii{}/H$\alpha$ ratio and gas-phase metallicity, as nitrogen is produced alongside oxygen in the CNO cycle.  The behavior of \sii{} is more complex.  Sulfur is produced via $\alpha$-capture, hence is created mostly in massive stars \citep[$>25 \mathcal{M}_{\odot}$;][]{weaver78, french80}, giving its abundance a weaker correlation with metallicity.  \sii{} emission luminosity specifically is a function of sulfur abundance and ionising radiation hardness \citep[primarily the latter's connection to gas temperature, as \sii{} is predominantly collisional;][]{osterbrock06}.  SII's ionisation potential is, however, very close to that of the much more abundant He I (23.3~eV vs. 24.6~eV), so hard ionising radiation preferentially ionises He I, while slightly softer radiation might preferentially ionise SII into SIII, decreasing \sii{} emission.

Fig.~\ref{fig:niisii} shows how the \nii{}/\sii{} ratio and the \sii{}/H$\alpha$ ratio vary with gas-phase metallicity \citep[using the \oiii{} and \nii{} empirical calibration, O3N2, from][]{marino13} in the MAD galaxies.  The red dashed lines show the running mean relations, while the red dotted lines show the running standard deviations about those means.  M101 shows a strong radial metallicity gradient \citep{croxall16, garner22}, but most \hii{} regions within M101 have direct-method metallicities between $8.1 < \log($O$/$H$) < 8.6$.  Hence, we derived a radial \nii{}/\sii{} correction in the DIG using the direct-method radial metallicity gradient provided by Eq.~10 of \citet{croxall16}, and we assume a constant value of \sii{}/H$\alpha = 0.47$, the average value among all DIG spaxels in the MAD galaxy sample with metallicities between $8.1 < \log($O$/$H$) < 8.6$.  Using a variable \sii{}/H$\alpha$ fraction derived from the median curve in the right panel of Fig.~\ref{fig:niisii} yields a negligible change on our corrected fluxes, hence we opt for the simplicity provided by the constant value.

The median correction we applied for \nii{} and \sii{} emission in the DIG in M101 is $\sim1.3$.  In NGC~5474, we assume a constant \nii{}$/$\sii{}$=0.192$, derived from M101's low-metallicity outer disk, and the same value of \sii{}$/$H$\alpha = 0.47$, for a correction factor of $\sim1.6$.

\subsection{\hii{} region interlopers}
\label{ssec:interlopers}

\begin{figure*}
    \centering
    \includegraphics[scale=1.0]{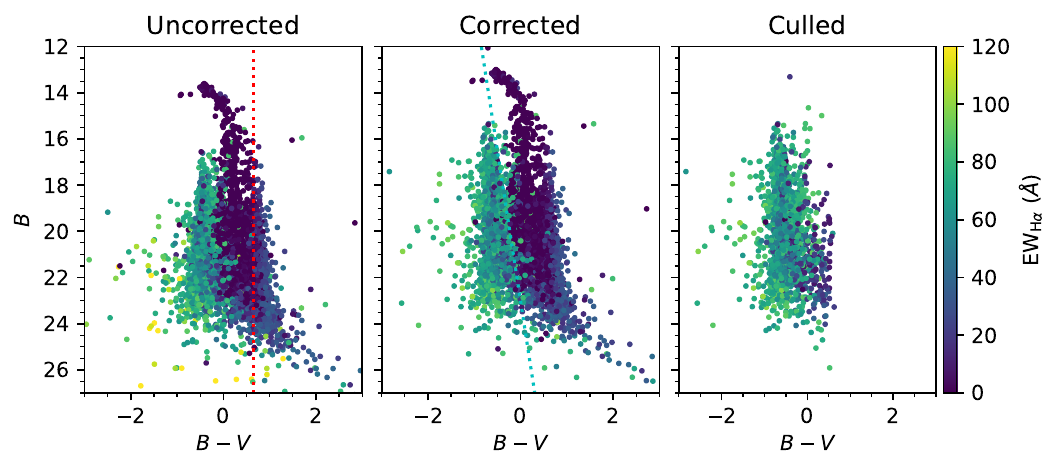}
    \caption{Demonstrating removal of interloping non-\hii{} region sources from our raw point-like source photometry catalogues.  The left panel shows the $B-V$ colour-magnitude diagram of all point-like sources detected in M101's vicinity, colour-coded by EW$_{{\rm H}\alpha}$, all corrected only for MW extinction.  The central panel shows photometry of the same sources, corrected for MW and internal extinction, with EW$_{{\rm H}\alpha}$ corrected for \nii{} and \sii{} contamination as well.  The right panel shows the sample culled of likely interloping sources.  The red dotted line in the left panel shows our initial colour-cut used to reject interloping MW stars ($B-V > 0.65$).  The dotted cyan line in the center panel shows the criteria we chose to identify \hii{} regions using the extinction- and line-contamination-corrected photometry.  We accepted as \hii{} regions all sources below the line, as well as any above the line not rejected as MW stars with EW$_{{\rm H}\alpha}>6$, necessary to preserve the faintest \hii{} regions in both galaxies.}
    \label{fig:cull}
\end{figure*}

To finalize our \hii{} region photometry catalogue, we needed to identify and remove interloping non-\hii{} region sources, typically foreground MW stars and background galaxies.  We did this using photometric cuts, demonstrated in Fig.~\ref{fig:cull}.

The left panel shows the $B-V$ colour-magnitude diagram (CMD) of all point-like sources we detected within and surrounding M101, colour-coded by H$\alpha$ EW in \AA.  We corrected $B$, $V$, and H$\alpha$ fluxes only for MW extinction in this panel.  The CMD is primarily composed of three types of objects: real \hii{} regions, with predominantly blue colours and high EW$_{{\rm H}\alpha}$; interloping MW stars, with a range of colours and predominantly low EW$_{{\rm H}\alpha}$, and interloping background galaxies, with a range of colours and a range of EWs.  As noted by \citet{watkins17} and \citet{garner21}, our H$\alpha$ filters are placed such that absorption features in low-temperature MW stars act to depress flux in our off-band filter relative to the continuum in our on-band filter, causing such stars to appear as detections with significant EW$_{{\rm H}\alpha}$ in our difference image\footnote{For example, the single point with EW$_{{\rm H}\alpha}\sim80$~\AA \ and $B-V\sim1.5$ is a nearby long-period variable star \citep[Tycho 3852-296-1;][]{hog00, gaia22}.}.  Most such stars are easily identifiable, however, by their red colours; they appear as the column of points with EW$_{{\rm H}\alpha} > 10$ and $B-V > 0.65$.  We thus removed anything with uncorrected $B-V>0.65$ from our \hii{} region catalogue.

The CMD of sources beyond $2\times R_{25}$ in either galaxy lacked the distribution of points with blue colours ($B-V \lesssim 0.5$) and high EW$_{{\rm H}\alpha}$ (EW$_{{\rm H}\alpha} \gtrsim 40$) seen in the left and central panels of Fig.~\ref{fig:cull}.  We thus applied a colour-cut to isolate \hii{} regions from stars, shown by the dotted cyan line in the central panel of Fig.~\ref{fig:cull}, which we make after applying extinction, \nii{}, and \sii{} corrections to all sources.  To balance star removal with preservation of faint \hii{} regions, we also excluded all such sources with corrected EW$_{{\rm H}\alpha}<6$.  The final \hii{} region catalogue CMD, corrected for all of the factors described in the preceding sections, is shown in the rightmost panel of Fig.~\ref{fig:cull}.  Our final catalogue contains 1954 \hii{} regions in M101 and 161 \hii{} regions in NGC~5474.

To assess the impact of our choice to limit \hii{} region photometry to only point-like emission, we estimated the high-luminosity power law index of M101's \hii{} region luminosity function.  The best-fit slope above $\log(L_{\hii{}})>37.5$ ergs s$^{-1}$ yields $\alpha = -1.98$, within the uncertainty of $\pm 0.2$ published by \citet{kennicutt89} for the same luminosity range.  This suggests that isolating flux measurements to only the brightest point-like parts of \hii{} regions is not unreasonable, and produces results consistent with other methods.  The largest differences likely arise for low flux objects, below the luminosity function knee.

\section{Results}
\label{sec:results}

\begin{table*}
\centering
\begin{tabular}{cccccccccccc}
\hline\hline
Relation & Galaxy & $\alpha$ & $\sigma_{\alpha}$ & $\beta$ & $\sigma_{\beta}$ & $\sigma_{\alpha,\,{\rm ext.}}$ & $\sigma_{\beta,\,{\rm ext.}}$ & $\sigma_{\alpha,\,{\rm NII,SII}}$ & $\sigma_{\beta,\,{\rm NII,SII}}$ & $\sigma_{\alpha,\,{\rm tot}}$ & $\sigma_{\beta,\,{\rm tot}}$ \\
$F_{\rm FUV}$ vs. $F_{{\rm H}\alpha}$ (HII) & M101 & 0.957 & 0.002 & 0.136 & 0.410 & 0.065 & 0.867 & 0.005 & 0.080 & 0.065 & 0.962 \\
$I_{\rm FUV}$ vs. $I_{{\rm H}\alpha}$ (DIG) & M101 & 1.293 & -- & 5.683 & -- & 0.191 & 3.140 & 0.018 & 0.302 & 0.192 & 3.154 \\
$Q_{10}$ vs. $I_{{\rm H}\alpha}$ & M101 & 0.735 & -- & -21.576 & -- & 0.000 & 0.000 & -- & 0.002 & -- & 0.002 \\
$F_{\rm FUV}$ vs. $F_{{\rm H}\alpha}$ (HII) & NGC 5474 & 1.099 & 0.046 & 2.623 & 11.576 & 0.146 & 2.142 & 0.005 & 0.115 & 0.153 & 11.773 \\
$I_{\rm FUV}$ vs. $I_{{\rm H}\alpha}$ (DIG) & NGC 5474 & 1.378 & -- & 7.393 & 0.023 & 0.616 & 10.694 & 0.124 & 2.185 & 0.628 & 10.915 \\
$Q_{10}$ vs. $I_{{\rm H}\alpha}$ & NGC 5474 & 0.992 & -- & -23.194 & 0.002 & 0.000 & 0.000 & -- & 0.006 & -- & 0.006 \\
\hline
\end{tabular}
\caption{Fit parameters on all regressions discussed in Sec.~\ref{sec:results}.  Fits are of the form $y = (\alpha \pm \sigma_{\alpha}) x + (\beta \pm \sigma_{\beta})$, where $x$ and $y$ are $\log_{10}$ of the parameters given in column 1.  Uncertainties on the fit parameters are given by the $\sigma$ columns.  The subscripts "ext" and "line" refer to systematic uncertainties on the fits induced by extinction corrections and corrections for \nii{} and \sii{} flux contamination, respectively, while $\sigma$ with no subscript refers to the uncertainty derived from the covariance matrix of the fit residuals.  We denote negligible uncertainties using --.  The final two columns provide the quadrature sum of all uncertainties for each fit parameter.}
\label{tab:fits}
\end{table*}

In the event that most DIG emission arises due to leaked LyC photons from \hii{} regions, we would expect a strong correlation between DIG H$\alpha$ surface brightness (hereafter, $I_{{\rm H}\alpha}$) and the luminosities of nearby \hii{} regions.  Likewise, if most DIG emission arises from in-situ field O and B stars, we would expect a strong correlation between DIG $I_{{\rm H}\alpha}$ and the cospatial FUV surface brightness ($I_{\rm FUV}$).  Hence, in this section, we showcase two correlations: that between DIG $I_{{\rm H}\alpha}$ and the incident ionising flux of each DIG pixel's nearest ten \hii{} regions, estimated from their H$\alpha$ luminosities; and that between DIG $I_{{\rm H}\alpha}$ and $I_{\rm FUV}$ of the same regions.  For comparison, we also show the relationship between H$\alpha$ and FUV flux among the point-like \hii{} regions.  Where appropriate, we also convert the H$\alpha$ and FUV surface brightnesses and fluxes into SFRs using the same calibrations as \citet{lee09} and \citet{byun21}, derived from \citet{kennicutt98} and \citet{kennicutt12}.  We concentrate on these two scenarios (leakage and field stars) because both M101 and NGC~5474 show prominent, wide-spread star formation.  We thus expect the contribution from HOLMES to be relatively small, and we lack the detailed spectroscopy necessary to identify shocked gas emission.

For each fit, we employed the {\sc scipy.odr} (V.1.10.0) implementation of orthogonal distance regression \citep[ODR;][]{boggs89}, weighting the fits on each axis by the combined photometric and calibration uncertainty where applicable.  We adopted calibration uncertainties of 10\% for both H$\alpha$ and FUV \citep{morrissey07, garner22}.  For H$\alpha$, the value derives from the combined uncertainty between the on- and off-band zeropoints, with a small amount added to consider uncertainty from photometric corrections (Sec.~\ref{ssec:corrections}).  This simply sets a ceiling to the weights; we find that the fits are insensitive to the exact values adopted, within reason.

We provide the best-fit slopes and intercepts for each relation we discuss in this section in Table~\ref{tab:fits}, including both standard errors derived from the covariance matrix of the regression residuals, and systematic uncertainties from our extinction, \sii{}, and \nii{} contamination corrections.  Where the uncertainties are negligible, as in the case of the standard errors when the number of points used in the fit is large (e.g., the $I_{{\rm H}\alpha}$--$I_{\rm FUV}$ relations, which use tens to hundreds of thousands of points), we do not report them.

We derived the systematic uncertainties using a Monte-Carlo approach, with $N=100$ iterations per fit.  First, we perturbed the best-fit parameters of the extinction, \nii{}, and \sii{} relations used to derive the corrections (radial for the former, metallicity for the latter) using normal distributions with $\mathcal{N}(0, \sigma_{\alpha})$ and $\mathcal{N}(0, \sigma_{\beta})$, where $\sigma_{\alpha}$ and $\sigma_{\beta}$ are the standard errors on the slopes and intercepts, respectively, of each systematic relation.  Additionally, we perturbed the mean \sii{}$/$H$\alpha$ ratio we applied for that correction by the standard deviation of that ratio in each case (\hii{} region and DIG).  We then re-derived all relevant fluxes using these perturbed relations, and re-derived each correlation using the same ODR approach as before.  We adopted the standard deviation of the $N=100$ perturbed fit parameters as the systematic uncertainty in each case.  We provide total uncertainties on the fit parameters in the last two columns of Table~\ref{tab:fits}, which are the quadrature sum of all uncertainties.

As one additional but important point, we measure the total H$\alpha$ fluxes within M101 and NGC~5474 to be $\log(F_{{\rm H}\alpha})=-10.00$ erg s$^{-1}$ cm$^{-2}$ and $-11.54$ erg s$^{-1}$ cm$^{-2}$, respectively (as this is merely a side-note, we eschew a formal estimate of uncertainties on these values, but at minimum they are $\pm 0.05$, or $\sim10$\% from the flux calibration).  These are very similar, within the uncertainties, to the values published by \citet{kennicutt08} of $-10.23\pm0.13$ and $-11.55\pm0.05$, respectively (adjusted for our adopted distance of 6.9~Mpc), in agreement with the conclusions of \citet{lee16}.  Interestingly, the largest discrepancy is with the more massive galaxy, M101, where it is not expected \citep{lee09}.

\subsection{DIG H$\alpha$ surface brightness vs. incident flux from \hii{} regions}
\label{ssec:q10}

\begin{figure*}
    \centering
    \includegraphics[scale=1.0]{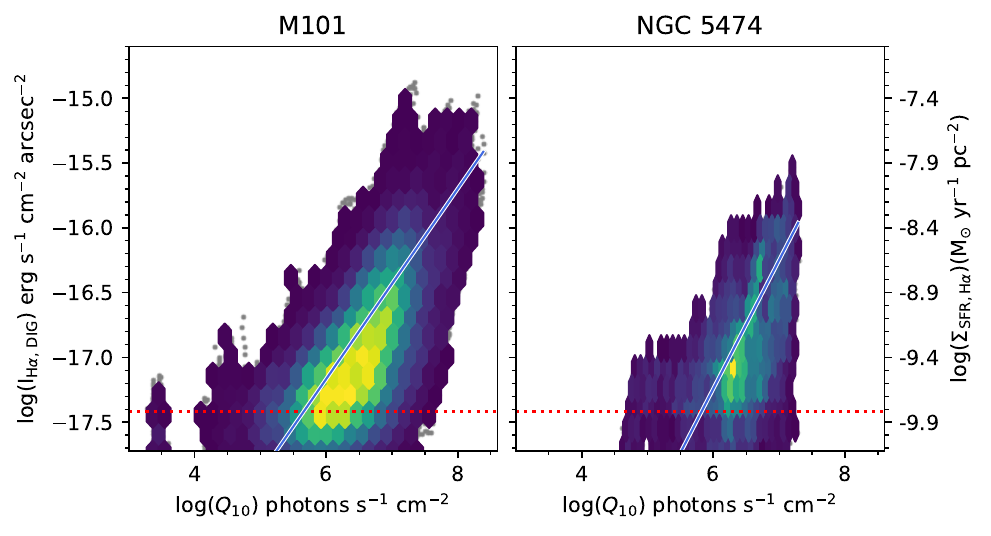}
    \caption{H$\alpha$ surface brightness in the DIG as a function of the number of ionising photons incident on each DIG pixel from the nearest 10 \hii{} regions, derived from the H$\alpha$ fluxes of said regions, uncorrected for internal extinction.  DIG pixels without significant H$\alpha$ flux are excluded here.  The colour scale in the 2D hexagon histogram denotes the density of points in each bin, with individual points shown in gray underneath.  The blue lines show the best-fit relations (Table~\ref{tab:fits}).  The dotted red line denotes the RMS in the H$\alpha$ image background, which sets our noise limit.}
    \label{fig:q10}
\end{figure*}

To consider the impact of \hii{} region LyC leakage on DIG, Fig.~\ref{fig:q10} shows the correlation between the $I_{{\rm H}\alpha}$ of the DIG and the incident ionising flux from the nearest ten \hii{} regions at each DIG pixel, for both M101 (left) and NGC~5474 (right).  We show the best-fit correlation for each galaxy as solid blue lines.  We chose the nearest ten regions because we found the contribution from any regions beyond this to be negligible, and changing the analysis to the nearest nine or eight regions made no substantive change to our results.

We estimated the incident ionising flux by summing the geometrically diluted H$\alpha$ flux of the nearest ten \hii{} regions incident on each DIG pixel, estimated from the regions' H$\alpha$ luminosities.  We converted this total H$\alpha$ flux to an ionising radiation flux using the relation $L($H$\alpha) = 1.37 \cdot 10^{-12} Q$ \citep{osterbrock06}, where $Q$ is the ionising photon rate in units of photons s$^{-1}$.  We denote this incident ionising flux as $Q_{10}$, which has units of photons s$^{-1}$ cm$^{-2}$.  We do not correct the \hii{} region luminosities for internal extinction in this step, as we assume that the incident flux at any given DIG pixel is that which escapes from the \hii{} region, not its geometrically diluted intrinsic brightness.  As such, systematic uncertainties on these values do not include any contribution from an extinction correction, although the relations change very little if we do employ extinction-corrected \hii{} region luminosities.

Both galaxies show a strong positive correlation between incident ionising flux and DIG $I_{{\rm H}\alpha}$, with slopes between $\sim 0.7$--$1$ (Table~\ref{tab:fits}), implying that leakage from \hii{} regions is an important contributor to the DIG in both galaxies.  We are, of course, making some physical assumptions by estimating the incident ionising flux in this manner.  If LyC photons leak from the \hii{} regions, the value of $Q$ we estimate for each \hii{} region from $L_{{\rm H}\alpha}$ corresponds to the total value of $Q$ minus the fraction which escapes into the ISM.  Thus, by assuming that the value of $Q$ we estimate for each \hii{} region is the same as that leaking out to ionise the DIG, we implicitly assume $f_{\rm esc}=0.5$.  We explore this assumption's validity in Sec.~\ref{ssec:digorigin}, in which we investigate how well our measured relation agrees with estimates from models, and what this implies about the fraction of DIG ionisation contributed by this leakage compared to the other potential source we investigate, field O and B stars.

\subsection{H$\alpha$ vs. FUV}
\label{ssec:fuvha}

\begin{figure*}
    \centering
    \includegraphics[scale=1.0]{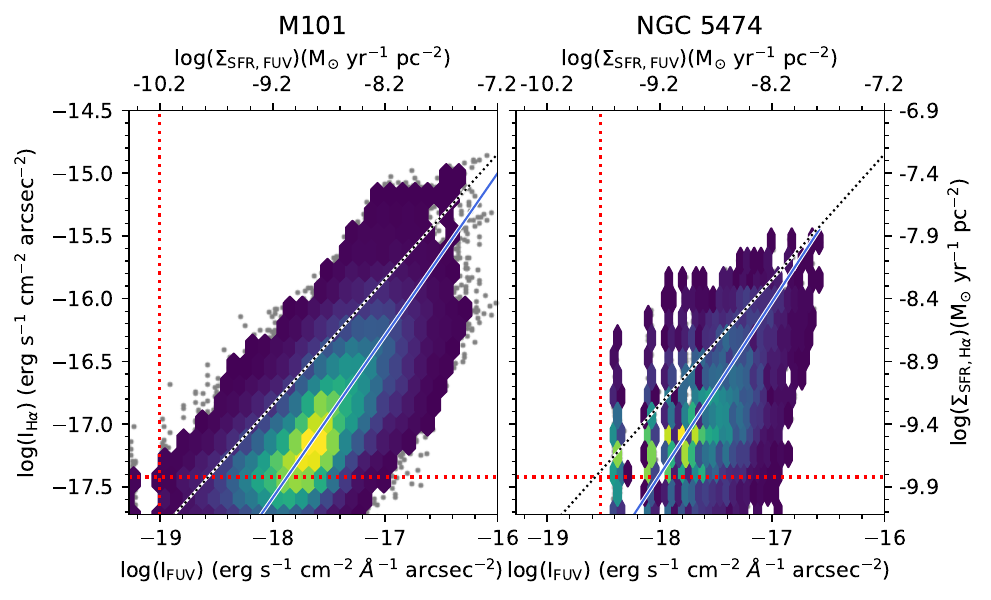}
    \caption{Pixel-to-pixel FUV vs. H$\alpha$ surface brightness in the DIG of both galaxies.  The plotting schema is the same as \ref{fig:q10}, but we have included also a vertical red dotted line showing the RMS in the FUV image backgrounds.  The top and right axes show surface brightness converted to SFR surface density from either band (FUV at the top and H$\alpha$ at the right).  Dotted black lines show the 1:1 relation in SFR.}
    \label{fig:hafuv_dig}
\end{figure*}

\begin{figure*}
    \centering
    \includegraphics[scale=1.0]{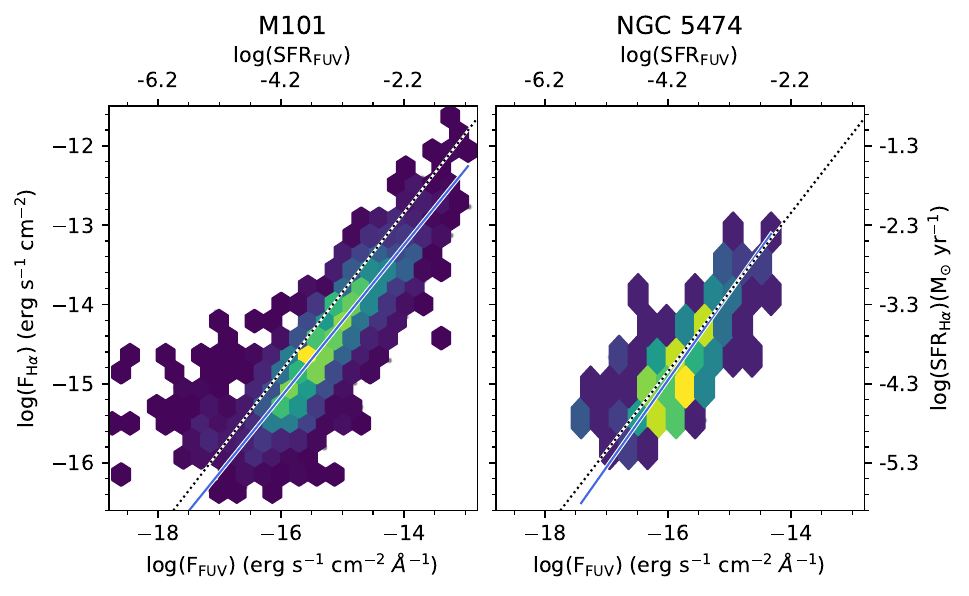}
    \caption{As Fig.~\ref{fig:hafuv_dig}, but for \hii{} regions.  The axis limits differ here, as the luminosities probed are much higher.  RMS limits fall outside of the axis limits in this figure.}
    \label{fig:hafuv_hii}
\end{figure*}

A strong correlation between diffuse H$\alpha$ and the FUV emission in the DIG regions may suggest that field O and B stars contribute heavily to the DIG emission, while a break in such a relation would suggest a transition from one DIG regime (perhaps dominated by \hii{} region leakage) to another.  To assess this potential contribution in the M101 Group, we show the correlation between H$\alpha$ and FUV surface brightness and flux for the DIG and \hii{} regions in Fig.~\ref{fig:hafuv_dig} and Fig.~\ref{fig:hafuv_hii}.  As before, we show the best-fit correlations as solid blue lines.  Dotted black lines show the relation where the SFR ratio is unity.  Dotted red lines in Fig.~\ref{fig:hafuv_dig} designate the RMS in the background, which serves as limiting surface brightnesses in both bands for the pixel-to-pixel photometry.

In the DIG (Fig.~\ref{fig:hafuv_dig}), we see a correlation with an unbroken slope close to one ($\sim 1.3 \pm 0.2$), with the SFR$_{\rm{H}\alpha}/$SFR$_{\rm FUV}$ ratio declining as a function of surface brightness in both bands.  This suggests that, while there may be a connection between the diffuse FUV and H$\alpha$ components in these galaxies, it is likely not as straight-forward as direct ionisation by field O and B stars.  We discuss such stars' possible contribution in Sec.~\ref{ssec:digorigin}.

In the \hii{} regions, the SFR ratio is constant across the full range of flux values.  In M101, this ratio is below unity (in linear units, SFR$_{\rm{H}\alpha}/$SFR$_{\rm FUV} \sim 0.44$).  In NGC~5474 it is consistent with unity, although the fit uncertainty is much higher than for M101.  Regardless, this near constant SFR ratio among \hii{} regions provides a seeming contrast to results from past studies \citep[e.g.,][]{lee09, lee16, byun21}.  We discuss the implications of this in Sec.~\ref{ssec:definitions}.

\section{Discussion}
\label{sec:discussion}

\begin{figure}
    \centering
    \includegraphics[scale=1.0]{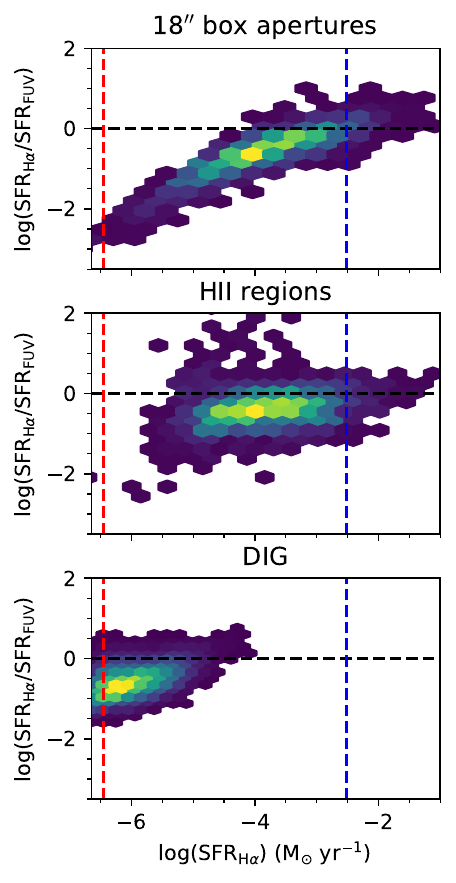}
    \caption{H$\alpha$- to FUV-derived SFR ratio as a function of H$\alpha$-derived SFRs measured in different environments.  The top panel shows these values measured using a grid of 18\arcsec$\times$18\arcsec \ box apertures, summing all flux (i.e., not H$\alpha$-selected only) in each box.  The central and bottom panels show these values for our point-like source \hii{} region and pixel-to-pixel DIG samples, respectively.  The black horizontal line denotes equality in both SFR indicators.  The blue vertical line denotes SFR$=0.003 \mathcal{M}_{\odot}$~yr$^{-1}$, the value below which \citet{lee09} and \citet{byun21} found the two SFR indicators to diverge.  The red vertical line denotes our limiting H$\alpha$ surface brightness, converted to SFR.}
    \label{fig:byuncomp}
\end{figure}

\begin{figure*}
    \centering
    \includegraphics[scale=1.0]{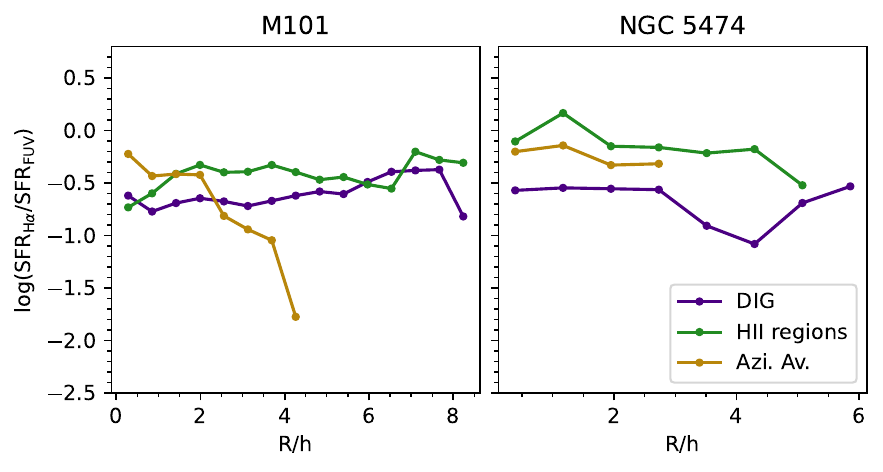}
    \caption{Azimuthally averaged radial profiles of SFR ratio, for three different cases: DIG pixels only (purple), \hii{} regions only (green), and all flux (gold), including FUV emission with no H$\alpha$ counterpart.  Gold profiles truncate where background begins to dominate in the H$\alpha$ image.  Radii are scaled by each galaxy's $B$-band scale length, which are 4.42~kpc \citep[2.2\arcmin;][]{mihos13} and 1.29~kpc (0.64\arcmin, measured for this work) for M101 and NGC~5474, respectively.}
    \label{fig:radprofs}
\end{figure*}

\subsection{Separation of \hii{} regions and DIG}
\label{ssec:definitions}

We showed in Sec.~\ref{sec:results} that the average trend among what we defined as DIG pixels showed depressed SFR$_{\rm{H}\alpha}/$SFR$_{\rm FUV}$ universally, declining with $I_{{\rm H}\alpha}$.  We also found that H$\alpha$ SFRs measured from the point-like objects which we identified as \hii{} regions are about half those predicted by the FUV fluxes at all luminosities.  So while the DIG trend seemingly reflects what was discovered in past investigations of this ratio \citep[e.g.,][]{meurer09, lee16, byun21}, the \hii{} region trend does not, despite that we use the same SFR calibrations as those studies.

\citet{lee09} found that the two SFR indicators diverge below SFR$_{{\rm H}\alpha} \sim 0.003 \mathcal{M}_{\odot}$~yr$^{-1}$ ($\log($SFR$_{{\rm H}\alpha}) \sim -2.5$) using integrated SFRs of dwarf galaxies.  \citet{byun21} later corroborated this finding locally within two spiral galaxies, measuring SFRs within 6\arcsec \ circular apertures ($\sim 300$~pc at their targets' distances) positioned on a hexagonal grid.  \citet{goddard10} also found that H$\alpha$ truncates much more rapidly than FUV in the azimuthally averaged surface brightness profiles of many disk galaxies.  Each of these studies thus used photometry of regions with larger spatial scales than our study, with apertures that likely contained both DIG and \hii{} regions.

Part of the discrepancy between our results and these others may thus be methodological, as we measure our fluxes through separation of point-like star-forming regions and diffuse regions.  We demonstrate this in Fig.~\ref{fig:byuncomp}, which shows $\log($SFR$_{\rm{H}\alpha}/$SFR$_{\rm FUV})$ as a function of $\log($SFR$_{\rm{H}\alpha})$ for three different cases.  In the top panel, we measured this ratio using a grid of box apertures across M101, with sizes of 18\arcsec$\times$18\arcsec \ \citep[$\sim600$~pc at M101's distance, to mimic][]{byun21}, summing all flux (\hii{} region, DIG, and FUV with no H$\alpha$ counterpart) within each box.  The bottom two panels show this same trend for our point-like \hii{} region and pixel-to-pixel DIG region samples, measured as described in Sec.~\ref{sec:methods}.  The horizontal black dashed lines show equal SFRs, while the vertical blue dashed lines show SFR$_{{\rm H}\alpha} \sim 0.003 \mathcal{M}_{\odot}$~yr$^{-1}$.  The red vertical dashed lines are our limiting H$\alpha$ surface brightness converted to a per-pixel SFR.

Using these larger box apertures, we do reproduce the trend found by \citet{byun21}, where only regions with the highest SFRs (near SFR$_{{\rm H}\alpha} \sim 0.003 \mathcal{M}_{\odot}$~yr$^{-1}$) show SFR ratios approaching unity.  However, the other two panels provide additional context: when summing both DIG and point-like \hii{} region flux, the trend appears as a kind of convolution of the flat \hii{} region trend and the declining DIG trend.

Fig.~\ref{fig:radprofs} also demonstrates this using azimuthally averaged profiles of $\log($SFR$_{\rm{H}\alpha}/$SFR$_{\rm FUV})$, shown as a function of disk scale length as measured in the $B$-band.  Each curve shows the average flux within concentric circular annulus apertures for three different cases: DIG pixels only (purple), \hii{} regions only (green), and all flux (gold; again, including FUV flux with no H$\alpha$ counterpart).  While the ratio remains fairly constant in both the DIG and \hii{} region curves (at $\sim 0.3$ and $\sim 0.5$ in linear units, respectively), the azimuthally averaged profile using all flux shows a strong decline in the ratio in M101 and a subtle decline in NGC~5474.  Beyond a few scale lengths in either galaxy, the azimuthally averaged H$\alpha$ flux has dropped to nearly zero, hence both profiles truncate.

The decline in the SFR ratio with SFR$_{{\rm H}\alpha}$ in the M101 Group thus seems to result from a transition from the regime spanned by the point-like \hii{} regions, where the ratio is constant, to the DIG regime, where it declines.  The reason for the decline in the DIG regime may result from a change in FUV-emitting stellar populations there compared to the bright young clusters found within the \hii{} regions.  Both M101 and NGC~5474 show abundant FUV emission with no H$\alpha$ counterpart, with it being particularly prevalent in M101's outer disk.  Indeed, we found that the fraction of pixels in each 18\arcsec$\times$18\arcsec \ box used in the top panel of Fig.~\ref{fig:byuncomp} with significant FUV emission (above the background RMS) but no significant H$\alpha$ emission (below the RMS) shows a strong negative correlation with SFR.  Using an alternative DIG map based on the FUV image, we found that around 27\% of this H$\alpha$-less diffuse FUV emission would have a detectable H$\alpha$ counterpart in our imaging were the ratio $F_{{\rm H}\alpha}/F_{\rm FUV}$ in the DIG the same as it is in the \hii{} regions.  This could occur either if DIG stars are less massive (thus producing fewer LyC photons) than those in the \hii{} regions, or if the DIG environment has a much higher $f_{\rm esc}$.  We explore this in the following section.

\subsection{DIG origins in the M101 Group}
\label{ssec:digorigin}

\begin{figure*}
    \centering
    \includegraphics[scale=1.0]{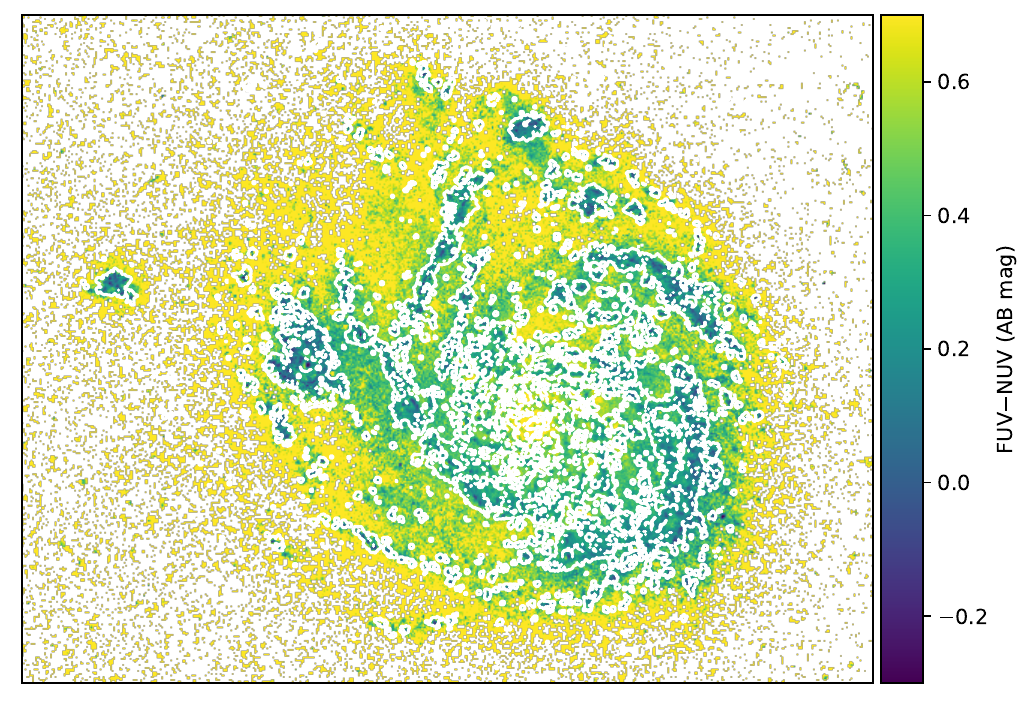}
    \caption{FUV$-$NUV colour map of M101, in AB magnitudes.  White contours outline the DIG and \hii{} regions.}
    \label{fig:uvcol}
\end{figure*}

\begin{figure}
    \centering
    \includegraphics[scale=1.0]{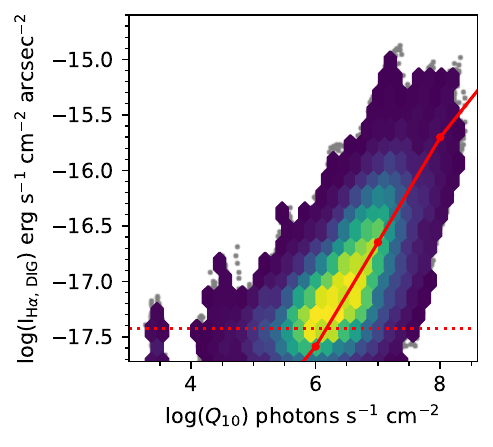}
    \caption{As the left panel of Fig.~\ref{fig:q10}, excluding the best-fit relation.  The red curve shows H$\alpha$ surface brightness predicted by a series of Cloudy simulations, in which the incident ionising flux on a plane-parallel cloud was set to values between $2 \leq \log(Q) \leq 10$ photons s$^{-1}$ cm$^{-2}$ (see text for details).}
    \label{fig:cloudy}
\end{figure}

As demonstrated in Sec.~\ref{ssec:fuvha}, there is a tight, nearly one-to-one correlation between H$\alpha$ and FUV surface brightness in the DIG.  This suggests that field O and B stars may contribute a substantial fraction of the power required to ionise it.  However, there is no break in the relation, as one might expect in LSB regions where ionisation from \hii{} regions has diluted.  We thus cannot rule out that the close correlation might arise simply because the two types of emission are tracing the same underlying phenomenon: that both young massive stars and DIG tend not to stray far from \hii{} regions, even if for very different reasons.

For example, \citet{oey18} estimated a velocity dispersion of field O and B stars (more massive than spectral type B0.5) in the Small Magellanic Cloud using Gaia Data Release 2 \citep{gaia18} of $\sim40$~km s$^{-1}$ in any one direction.  If this is comparable in M101, over a 10~Myr timespan (roughly the lifespan of such stars), these would travel $\sim400$~pc from their natal clusters on average.  For comparison, the median distance between all DIG pixels and their nearest-neighbour \hii{} regions in M101 is 389~pc, and in NGC~5474 is 358~pc.

DIG thus does not stray much farther from \hii{} regions than typical field O and B stars would if those stars originated within the same regions as the current star formation.  DIG also exists primarily in regions with high gas density, where on-going star-formation is more likely.  By cross-matching our DIG pixel coordinates with the \hi{} moment zero map of M101 from The \hi{} Nearby Galaxy Survey \citep[THINGS;][]{walter08}, we found that the \hi{} emission in DIG pixels shows a fairly steady value of $\log(N_{\hi{}}) = 20.67$~cm$^{-2}$ ($\sim 3.4 \mathcal{M}_{\odot}$~pc$^{-2}$) across either galaxy, fairly typical of star-forming regions in spirals \citep[e.g.,][]{bigiel08}.

One way in which we can assess the likelihood that these field O and B stars are powering the DIG is by examining the diffuse FUV stellar populations through integrated colour.  Fig.~\ref{fig:uvcol} shows an FUV$-$NUV (AB magnitudes) colour map of M101, with white contours outlining the DIG and \hii{} regions.  Here, we have corrected both FUV and NUV flux for extinction as described in Sec.~\ref{ssec:corrections}.

It is clear at a glance that FUV-emitting populations located outside of either DIG or \hii{} regions show systematically redder colours than those located within those regions, and that \hii{} regions themselves show bluer colours than DIG regions.  To be more quantitative, the average colour within the white contours is FUV$-$NUV$=0.31 \pm 0.24$, compared to FUV$-$NUV$=0.44 \pm 0.25$ outside of the contours (including all pixels, inner and outer disk alike), while the \hii{} regions have a mean colour of $0.05 \pm 0.27$.  Most of the scatter in these colours seems to arise from variability in extinction rather than intrinsic variability or photometric uncertainty.  We reproduce the distribution of both DIG and \hii{} region colours well using their median colours perturbed by normally distributed extinction corrections with a standard deviation of 0.6~mag, roughly the scatter about our best-fit extinction gradient in M101.

Using the population synthesis software Code Investigating GALaxy Emission \citep[CIGALE;][]{burgarella05, noll09, boquien19}\footnote{\url{https://cigale.lam.fr/}}, we found that a colour as red as FUV$-$NUV$=0.3$ is difficult to produce in the presence of a substantial population of young O stars.  A population modelled as a recent (25~Myr ago) burst atop a constant SFR (a reasonable model of M101) always maintains colours $<0$, while a single fading starburst does not reach a colour of 0.3 until $\sim150$~Myr of age, by which point its ionising flux is too low by several orders of magnitude to produce even the lowest values of $I_{{\rm H}\alpha}$ we measure in the DIG.  Similarly, a simulation of a fading burst using the galaxy evolution software GALEV \citep{kotulla09} reaches the same colour by $\sim400$~Myr of age, by which time its ionising flux is vanishingly small.

This therefore suggests that, despite the spatial coincidence between DIG and diffuse FUV near \hii{} regions, the field O and B star contribution to DIG is minimal in the M101 Group, on average.  The large-scale diffuse FUV component in both galaxies could well be a remnant of the tidal interaction between M101 and NGC~5474 $\sim300$---$400$~Myr ago \citep{mihos13, mihos18, linden22}, with the FUV most coincident with the DIG being remnants of dissolved clusters from earlier episodes of star-formation in the spiral arms \citep[likely streamed there after forming within spiral arms; e.g.][]{crocker15, garner24}.  If the large-scale FUV emitted by these redder stars has a corresponding DIG equivalent, it lies at surface brightnesses below our sensitivity, and hence cannot be constrained using our data.

If field O and B stars contribute little, this diffuse gas must comprise a distinct physical environment from the point-like \hii{} regions.  We must therefore consider some alternative sources of ionisation.  A study by \citet{lacerda18} found that in Sc galaxies like M101, the LyC contribution from older, harder ionising sources such as HOLMES should be fairly small, assuming emission with EW$_{{\rm H}\alpha} < 3$\AA \ arises primarily from such sources.  The distribution of EW$_{{\rm H}\alpha}$ in M101 and NGC~5474 agrees with this, with only $\sim 10$\%\footnote{This is an upper limit.  In our images, EW$_{{\rm H}\alpha}$ declines with distance from \hii{} regions, suggesting low-EW$_{{\rm H}\alpha}$ regions could be the result of geometric dilution, not necessarily a change of ionisation source.} of the total H$\alpha$ flux in the DIG arising from pixels with EW$_{{\rm H}\alpha} < 3$\AA \ in either galaxy.  Using their criteria, the remainder of the DIG must be ionised by a mixture of sources, including photoionisation.

We thus turn our attention to leakage of LyC photons from \hii{} regions.  To assess the contribution from this source, we performed an array of simulations using the spectral synthesis code Cloudy \citep[Ver.~17.03;][]{ferland17} in an attempt to reproduce the trend between $\log(Q_{10})$ and DIG $I_{{\rm H}\alpha}$ displayed in Fig.~\ref{fig:q10}.  We created a synthetic young star cluster as our illumination source using the code Starburst99 \citep[Ver.7.0.0;][]{leitherer99, vazquez05, leitherer10, leitherer14}, with stellar mass of $10^{6} \mathcal{M}_{\odot}$.  With this illumination source, selecting an age of 2~Myr, we ran an array of Cloudy simulations using the $\Phi$(H) parameter option, which allows one to specify directly the incident ionising photon flux (in photons s$^{-1}$ cm$^{-2}$) on the surface of a cloud.  We used a cloud with a gas density of 1 cm$^{-3}$ as the target, and set $\log(\Phi$(H)$)=2$---$10$ in steps of $1$.  We then recorded the resulting cloud's emergent $I_{{\rm H}\alpha}$, excluding reflection and transmission as we are viewing these clouds in M101 and NGC~5474 from above, while the \hii{} region flux incident on the clouds would be predominantly in the disk plane.

In Fig.~\ref{fig:cloudy}, we overplot these model surface brightnesses on the $\log(Q_{10})$---$\log(I_{{\rm H}\alpha})$ relation from the left panel of Fig.~\ref{fig:q10}, as a red curve.  We found that the shape of this curve is insensitive to the ionising source, as, unlike line ratios, H$\alpha$ emission measure is merely a function of the ionisation and recombination rate and hence the total incident ionising flux, not the ionising spectrum \citep{field75}.  The curve shape is also insensitive to the chosen gas density, save for densities much higher than typically found in the DIG ($> 100$~cm$^{-3}$), and to the choice of filling factor and grain composition.

The match between the predicted and observed $I_{{\rm H}\alpha}$ is remarkable.  Above the image noise threshold, the close agreement between the predicted and observed values implies that the majority of the ionising flux producing DIG in M101 and its companion arises from LyC leakage from \hii{} regions.  Excluding the hard ionised DIG using the criteria from \citet{lacerda18}, this contribution would be $\gtrsim 90$\%.

As discussed in Sec.~\ref{ssec:q10}, we used the measured H$\alpha$ luminosities of each \hii{} region, diluted only geometrically within the disk plane, to estimate the LyC flux incident on each DIG pixel.  This presumes $f_{\rm esc}=50$\%, and the good agreement between the Cloudy models and our data provides support that this value is approximately correct.  Estimates from the literature seem to concur, albeit with a wide variability.  For example, a study by \citet{teh23} found that \hii{} regions with $\log(L_{{\rm H}\alpha}) < 38.06$ ergs s$^{-1}$ have $f_{\rm esc} \sim 0.56^{+0.08}_{-0.14}$ (lower for brighter populations).  In the M101 Group, $\log(L_{{\rm H}\alpha}) = 38.06$ ergs s$^{-1}$ is $\log(F_{{\rm H}\alpha}) \sim -13.7$ ergs s$^{-1}$ cm$^{-2}$; only $\sim15$\% of the \hii{} regions in M101 lie above this value.  In more tentative agreement, \citet{dellabruna21} find a value of $f_{\rm esc}\sim0.67^{+0.08}_{-0.12}$ among a small sample of mostly more luminous \hii{} regions in NGC~7793.  \citet{pellegrini12} likewise estimated a luminosity-weighted mean $f_{\rm esc} \sim 40$\% in the Large and Small Magellanic Clouds.

Molecular cloud evolution models show that $f_{\rm esc}$ is a strong function of age, rising from nearly zero to nearly one within around 5~Myr \citep[depending on the ionising cluster's luminosity, the gas-phase metallicity, the star formation efficiency, and other factors; e.g.][]{rahner17, kimm22}.  If so, it may not be surprising if the \hii{} region population in a galaxy with a steady SFR over gigayear timescales has a mean $f_{\rm esc}$ falling roughly halfway between zero and one.  This does not mean that all radiation escaping from the \hii{} regions escapes the galaxy as a whole, of course: likely this escape is not omni-directional, and whether this escaping emission ionises the galaxy's own ISM or leaves to ionise the IGM depends both on the directionality of the escape from the cloud and the location of the cloud within the galaxy itself \citep[e.g.,][]{pellegrini12, kim23}.

Because SFR and LyC flux are both related to $L_{{\rm H}\alpha}$ by a scale factor, a loss of half of the LyC flux to leakage would reduce the SFR$_{{\rm H}\alpha}$ estimated from $L_{{\rm H}\alpha}$ by a factor of two.  Assuming the values of SFR$_{\rm FUV}$ are accurate, this agrees well with the average value of SFR$_{{\rm H}\alpha}/$SFR$_{\rm FUV} \sim 0.44$ we find among the \hii{} regions.  In one way, this is expected: the SFR calibration we employ was initially derived by \citet{kennicutt83}, who in the subsequent iterations we employ \citep{kennicutt98, kennicutt12} still assumed a constant SFR over $>100$~Myr timescales and Case B recombination \citep{brocklehurst71} without LyC leakage, which they state provides a lower limit on the true SFR.  Even so, estimating the LyC leakage fraction is not a trivial exercise, so the value we propose here, while reasonable, should be considered a fairly rough estimate.

This experiment is an alternative version of that performed by \citet{zurita02}, \citet{seon09}, and most recently \citet{belfiore22}, who used the measured \hii{} region fluxes in their galaxies to predict the DIG surface brightness distribution.  Our model differs in that we did not include attenuation of the ionising radiation through the interstellar medium.  However, our results mirror theirs insofar as they imply a very large mean free path ($>1$~kpc) is required to explain the DIG surface brightness via \hii{} region leakage alone.  \citet{seon09} claimed that the necessary absorption coefficient was unphysically low for reasonable models of the ISM, however \citet{belfiore22} explained this by suggesting it may result from DIG lying preferentially above the cold gas disk (with scale height $\sim 100$~pc), in a region where most of the ISM is ionised \citep[in accord with studies of DIG in edge-on galaxies, where DIG scale heights are of order 1--2 kpc;][]{collins01, miller03, levy19, rautio22}.

As M101 and NGC~5474 are both face-on, we cannot easily corroborate this explanation, but the good agreement between our simple Cloudy model comparison with our data and the results of these past studies provides further evidence that leakage of LyC photons from \hii{} regions is sufficient to explain most of the ionising power of the DIG here.  The discrepancy between integrated H$\alpha$ and FUV SFRs in the LSB regime in this group may therefore reflect the combination of a longer SFR duty cycle in that regime, leading to a more noticeable mixture of old and young massive stars, and the tendency for \hii{} regions to lose $\sim50$\% of their LyC photons to leakage, on average.  The former may be a direct consequence of the group's unique interaction history, so it is unclear how transferable this might be to other LSB environments.

\subsection{Implications beyond the M101 Group}

Having established that what we have defined, morphologically, as DIG represents a distinct physical environment compared to the point-like sources we identified as \hii{} regions, we consider here a unifying physical model of the ionised gas in M101 and its companion.  We do this by relating our observations to those of H$\alpha$ emission in the MW.

\citet{madsen06} found that classical \hii{} regions in our Galaxy (bright emission line regions immediately surrounding hot stars) show much more consistent temperatures and line ratios than DIG (everything else).  DIG temperature, by contrast, appears to depend on its distance from such regions, or else it depends on the specific ionisation mechanism (e.g., supernova feedback rather than photoionisation).  In our scenario, where DIG seems primarily ionised by LyC leakage, the gradually degrading relationship between SFR$_{{\rm H}\alpha}$ and SFR$_{\rm FUV}$ might be illustrating similar behavior in M101 and its companion.

One obvious problem with our DIG definition is its dependence on image resolution, however.  For example, even though M101 is nearby, many of its faintest \hii{} regions, with small diameters, could blend in with what we defined as DIG, imposing a hard, resolution-dependent size cutoff on what we define as \hii{} regions.  Also, \hii{} regions at advanced ages tend to be more diffuse and patchy than younger regions \citep{hannon19}, which would also blend in with DIG in unresolved imaging, yielding an age-limit on this \hii{} region definition as well \citep[assuming that evolved \hii{} regions are fundamentally distinguishable from DIG, which they may not be;][]{rousseau18}.  More distant galaxies would suffer more from these systematic effects, and their impact on the correlations we explore here would be difficult to discern without a comparison study using higher resolution imaging of the same regions.  Even so, our analysis does suffer less from such effects compared to those using integrated H$\alpha$ and FUV fluxes from whole galaxies, or even those using integrated fluxes over significantly larger apertures than what we use here ($> 100$~pc).

If what we observe in the M101 Group is universal, DIG is comprised primarily of gas ionised by leakage from \hii{} regions, and so total H$\alpha$ emission should constitute an accurate estimate of the instantaneous SFR \citep[e.g.,][]{magana20}.  In low-SFR environments, however, such as dwarf galaxies, LSB galaxies, and outer disks, where the classical \hii{} region density is low \citep[e.g.,][]{schombert13}, the instantaneous SFR and the longer-term SFR probed by FUV emission would tend to diverge as observed, depending on the relative extent of FUV compared to H$\alpha$ emission.  If the IMF is invariant \citep[as suggested, in the M101 Group, by our previous results;][]{watkins17}, this scenario suggests that variability in the SFR$_{{\rm H}\alpha}/$SFR$_{\rm FUV}$ ratio in the LSB regime is purely an artifact of the longer star formation duty cycle in that regime.  If the IMF is variable, however, an idea with some observational and theoretical support \citep[e.g.,][and many others]{meurer09, pflamm09, conroy12, geha13, li23}, the variability in that ratio must arise from a complex interaction between that IMF variability and the star-formation duty cycle.  Extrapolating our methodology to other nearby galaxies may help to disentangle these competing scenarios.

\section{Summary}
\label{sec:summary}

We present H$\alpha$ and FUV photometry of diffuse ionised gas (DIG) and \hii{} regions in the nearby M101 Group.  We find a strong correlation between the H$\alpha$ surface brightness ($I_{{\rm H}\alpha}$) in the DIG and the incident ionising flux on each DIG region from its nearest ten \hii{} regions ($Q_{10}$), assuming a Lyman continuum escape fraction of $f_{\rm esc}=0.5$.  This suggests that flux leakage from \hii{} regions is an important contributor to DIG.  Likewise, we find a strong correlation between H$\alpha$ and FUV surface brightness in DIG regions, suggesting that field O and B stars may contribute as well.

However, integrated FUV$-$NUV colours of DIG regions are quite red ($\sim 0.3$) compared to \hii{} regions ($\sim 0.05$), implying the young stellar populations embedded within the DIG are predominantly low-mass and thus likely contribute little to DIG ionisation.  By contrast, using a suite of Cloudy models, in which we ionize a slab of gas with a slew of ionising photon fluxes, we reproduced the correlation between $I_{{\rm H}\alpha}$ and $Q_{10}$ very well.  This suggests that most of the DIG in the M101 Group can be explained by leakage of Lyman continuum photons from \hii{} regions, with little contribution from field OB stars or other sources.  The excellent match between this predicted and observed $I_{{\rm H}\alpha}$--$Q_{10}$ relation is intriguing, as it implies the value of $f_{\rm esc}=0.5$ we chose is correct \citep[in tentative agreement with past studies; e.g.][]{dellabruna21, teh23}.  Also, as we did not include absorption within the interstellar medium (ISM) in our models, we find good agreement with similar analyses of other galaxies, in which the estimated mean free path of ionising radiation in the ISM is very high \citep[$>1$~kpc;][]{seon09, belfiore22}.

We compared the star formation rates (SFRs) derived from H$\alpha$ and FUV in both the DIG and \hii{} regions.  In the DIG, H$\alpha$ under-predicts SFR compared to FUV everywhere, with the ratio SFR$_{{\rm H}\alpha}/$SFR$_{\rm FUV}$ declining as a function of SFR$_{{\rm H}\alpha}$.  In the \hii{} regions, which we define as all point-like sources identified within both galaxies from our H$\alpha$ difference image, the ratio is flat at a value of $0.44$ to the faintest regions we detect (SFR$_{{\rm H}\alpha}\sim10^{-5}\mathcal{M}_{\odot}$~pc$^{-2}$).  Given this, we suspect that these point-like regions are mostly leaky, compact Str\"{o}mgren spheres, while the DIG comprises a mix of faint or old \hii{} regions and true diffuse gas.

By doing photometry within larger apertures which mix DIG, \hii{} regions, and FUV with no H$\alpha$ counterpart---both using boxes with widths of $\sim 300$~pc and using azimuthal averaging---we reproduce a trend found in other galaxies, in which SFR$_{{\rm H}\alpha}/$SFR$_{\rm FUV}$ decreases with SFR$_{{\rm H}\alpha}$ below SFR$_{{\rm H}\alpha}\sim0.003 \mathcal{M}_{\odot}$~pc$^{-2}$ \citep{lee09, lee16, byun21}.  Diffuse FUV without a detectable H$\alpha$ counterpart is wide-spread throughout the M101 Group, mostly in regions with low surface brightness, which explains why including all flux in the apertures (not only H$\alpha$-selected flux) results in this trend here.

The M101 Group's star formation history is defined by a recent interaction between M101 and NGC~5474, resulting in a burst of star formation 300--400~Myr ago \citep{mihos13, mihos18, linden22}.  Thus, the declining SFR ratio with SFR$_{{\rm H}\alpha}$ we find on these large scales in this group may result simply by mixing remnants from this burst (detectable as FUV emission with no H$\alpha$ counterpart) with on-going star formation (detectable as \hii{} regions and the diffuse gas they ionise around them).  Repeating this analysis---separating DIG from point-like, classical \hii{} regions---in other galaxies should help discern whether or not this is unique to the M101 Group.

\section*{Acknowledgements}

AEW acknowledges support from the STFC [grant numbers ST/S00615X/1 and ST/X001318/1].  JCM thanks the Mt Cuba Astronomical Foundation for funding support.  We thank the anonymous referee for their careful reading and thoughtful comments, which helped improve the quality of the manuscript.  We thank Johan Knapen, Chris Collins, and Sugata Kaviraj for many insightful discussions which helped shape the course of this work.  This work made use of Astropy:\footnote{http://www.astropy.org} a community-developed core Python package and an ecosystem of tools and resources for astronomy \citep{astropy:2013, astropy:2018, astropy:2022}.  This work also made use of the SciPy (ver. 1.10.0) library \citep{scipy20} and the NumPy (ver. 1.23.5) library \citep{numpy20}.  Figures in this paper were created using the Matplotlib (ver. 3.6.2) library \citep{matplotlib07}.  Calculations in Sec.\ref{ssec:digorigin} were performed with version 17.03 of Cloudy \citep{ferland17}. The authors are honored to be permitted to conduct astronomical research on Iolkam Du’ag (Kitt Peak), a mountain with particular significance to the Tohono O’odham.

\section*{Data Availability}

All GALEX data used in this study is available publicly through the Mikulski Archive for Space Telescopes (MAST), at the following URL: \url{https://galex.stsci.edu/GR6/}.  The Burrell-Schmidt Telescope data used in this study is available on reasonable request to the authors.



\bibliographystyle{mnras}
\bibliography{refs}






\bsp	
\label{lastpage}
\end{document}